
\magnification\magstep1
\scrollmode

\font\eightrm=cmr8                
\font\tensmc=cmcsc10              

\font\tenbbb=msbm10 \font\sevenbbb=msbm7 \font\fivebbb=msbm5
\newfam\bbbfam
\textfont\bbbfam=\tenbbb \scriptfont\bbbfam=\sevenbbb
    \scriptscriptfont\bbbfam=\fivebbb

\font\tengoth=eufm10 \font\sevengoth=eufm7 \font\fivegoth=eufm5
\newfam\gothfam
\textfont\gothfam=\tengoth \scriptfont\gothfam=\sevengoth
    \scriptscriptfont\gothfam=\fivegoth
\def\goth{\fam\gothfam}    

\def\opname#1{\mathop{\rm#1}\nolimits} 

\def\a{\alpha}                    
\def\A{{\cal A}}                  
\def \AU{{\cal A(U)}}             
\def\abs#1{\vert#1\vert}          
\def\B{{\cal B}}                  
\def\b{\beta}                     
\def\Bar#1{\overline{#1}}         
\def\bbuildrel#1_#2^#3{\mathrel{
       \mathop{\kern 1pt#1}\limits_{#2}^{#3}}}
\def\bw{\bigwedge\nolimits}       
\def\cite#1{\lbrack{\bf#1}\rbrack} 
\def\Coo{C^\infty}                
\def\cor#1#2{\lbrace#1,#2\rbrace} 
\def\de{\delta}                   
\def\De{\Delta}                   
\def\Der{\opname{Der}}            
\def\dim{\opname{dim}}            
\def\Ec{{\cal E}}                 
\def\ENE{{\cal N}}                
\def\eq#1{{\rm(#1)}}              
\def\Ga{\Gamma}                   
\def\harr#1#2{\smash{\mathop{\hbox to .5in{\rightarrowfill}}
       \limits^{\scriptstyle#1}_{\scriptstyle#2}}}
\def\hharr#1#2{\smash{\mathop{\hbox to .3in{\rightarrowfill}}
       \limits^{\scriptstyle#1}_{\scriptstyle#2}}}
\def\harrt#1#2{\smash{\mathop{\overline{\hbox to .5in{\hrulefill}}}
       \limits^{\scriptstyle#1}_{\scriptstyle#2}}}
\def\Hat#1{\widehat{#1}}          
\def\Hom{\opname{Hom}}            
\def\ID{\relax{1\kern-.24 em\rm l}}  
\def\Im{\opname{Im}}              
\def\L{{\cal L}}                  
\def\ker{\opname{ker}}            
\def\M{{\cal M}}                  
\def\Mg{{\goth M}}                
\def\o{\omega}                    
\def\om{\omega}                   
\def\Om{\Omega}                   
\def\ox{\otimes}                  
\def\o+{\oplus}                   
\def\O+{\bigoplus}                
\def\pd#1{{{\partial_{#1}}}}      
\def\pdf#1#2{{{\partial#1}\over{\partial#2}}}  
\def\pr{\opname{pr}}              
\def\Proof{\noindent{\sl Proof.\ \ }} 
\def\sepword#1{\qquad\hbox{#1}\quad} 
\def\set#1{\{\,#1\,\}}            
\def\si{\sigma}                   
\def\Si{\Sigma}                   
\def\smc{\tensmc}                 
\def\sopd#1#2#3{{{{\partial^2{#1}}\over{{\partial{#2}\partial{#3}}}}}}
\def\srule#1#2{(-1)^{\vert#1\vert\,\vert#2\vert}}

\def\stroke{\mathbin|}            
\def\th{\theta}                   
\def\Th{\Theta}                   
\def\T{{\bf T}}                   
\def\Tau{{\cal T}}                
\def\U{{\cal U}}                  
\def\Vker{\opname{Vker}}          
\def\varr#1#2{\llap{$\scriptstyle #1$}\left\downarrow
       \vcenter to .5in{}\right.\rlap{$\scriptstyle #2$}}

\def\w{\wedge}                    
\def\x{\times}                    
\def\X{{\goth X}}                 
\def\Y{{\cal Y}}                  
\def\z{\zeta}                     
\def\1{\'{\i}}                    
\def\3{\sharp}                    
\def\7{\dagger}                   
\def\.{\cdot}                     
\def\:{\colon}                    
\def\<#1,#2>{\langle#1\stroke#2\rangle} 

\def\qed{\allowbreak\qquad\null
           \nobreak\hfill\square}   
\def\square{\vbox{\hrule
               \hbox{\vrule height 5.2pt
                \hskip 5.2pt
                \vrule}\hrule}}     

\def\today{\number\day\space      
             \ifcase\month\or
              January\or February\or
              March\or April\or May\or
              June\or July\or August\or
              September\or October\or
              November\or December\fi
              \number\year}

\newcount\secnum
\outer\def\beginsection#1. #2\par{
        \vskip 0pt plus.1\vsize        
        \penalty -250                  
        \vskip 0pt plus-.1\vsize       
        \bigskip \vskip\parskip        
        \global\secnum=#1%
        \global\equationnum=0
        \message{#1. #2}%
        \leftline{\bf#1. #2}\nobreak
        \smallskip\noindent}

\outer\def\subsection#1.#2. #3\par{%
        \ifnum#2=1 \smallskip       
         \else \bigskip \fi         
        \message{Sec #1.#2.}%
        \leftline{\it#1.#2. #3}\nobreak
        \smallskip\noindent}

\def\declare#1. #2\par{\medskip   
            \noindent{\bf#1.}\rm    
            \enspace\ignorespaces
            #2\par\smallskip}


\newcount\equationnum             
\def\ecnum{\the\secnum.
             \the\equationnum}      
\everydisplay={\global\advance    
                 \equationnum by 1  
                 \leftnumdisplay}   
\def\leftnumdisplay#1{\let\nxxt#1\maybealign}
\def\maybealign{\ifx\nxxt\leqalignno
                  \else\leftnumeqn\fi\nxxt}
\def\leftnumeqn#1$${#1\leqno(\ecnum)$$}  

\def\eqlabel#1{\xdef#1{\ecnum}}   

\def\refno#1. #2\par{\smallskip   
            \item{\lbrack#1\rbrack}
                  #2\par}

\long\def\suspend#1\resume{}      

\def\batlleetal{1}
\def\berg{2}
\def\CarinenaFer{3}
\def\Hector{4}
\def\Hectoriii{5}
\def\Hectoriv{6}
\def\Hectorv{7}
\def\CarinenaLopez{8}
\def\CarinenaLoMari{9}
\def\CarinenaLoMarii{10}
\def\CarinenaLoRo{11}
\def\CrampinPirani{12}
\def\dirac{13}
\def\catalanes{14}
\def\tesis{15}
\def\GraciaPons{16}
\def\GraciaPonsdos{17}
\def\IbortMarin{18}
\def\kamimura{19}
\def\Kostant{20}
\def\Leites{21}
\def\Manin{22}
\def\MunozRoman{23}
\def\PugliVino{24}
\def\Sanchez{25}
\def\Skinner{26}
\def\SkinnerRusk{27}
\def\Vaintrob{28}


\centerline{\bf Singular Lagrangians in supermechanics}

\bigskip

\centerline{\smc Jos\'e F. Cari\~nena}

\smallskip

\centerline{\it Departamento de F\'{\i}sica Te\'orica,
Universidad de Zaragoza, 50009 Zaragoza, Spain.}

\medskip

\centerline{\smc H\'ector Figueroa}

\smallskip

\centerline{\it Escuela\ de Matem\'atica, Universidad de
Costa Rica, San Jos\'e, Costa Rica.}

\bigskip

\centerline{\bf Abstract}

\bigskip

{\narrower\eightrm\baselineskip=9.5pt       
\noindent

The time evolution operator $K$ is introduced in the graded context and
its main properties are discussed. In particular, the operator $K$ is
used to analize the projectability of constraint functions arising
in the Lagrangian formalism for singular Lagrangians.

\par}

\bigskip

\noindent{\it Keywords\/}:
time evolution operator, supervector fields along a morphism,
Hamiltonian and Lagrangian constraints.

\noindent{\it 1991 MSC numbers\/}:
Primary: 58A50, 58C50. Secondary: 70H33

\noindent{\it PACS numbers\/}:  03.20+i, 02.90+p, 11.30.pb.
\bigskip

\beginsection 1. Introduction

\medskip

The relevance of systems defined by singular Lagrangians for
fundamental physical theories
(generally  covariant, Yang Mills and string theories) is nowadays fully
understood. They are the only possibility  for the occurrence of gauge freedom.
Constraints, gauge invariance, gauge fixing, etc,
are now concepts of common use in these theories. All of them are
better understood
when using an appropriate geometric framework, and the use of modern
tools of Differential
Geometry has very much clarified the different aspects of the theory
of singular systems
started by Bergmann
\cite{\berg} and Dirac \cite{\dirac}.

The connection between the Lagrangian and Hamiltonian formalisms of
regular systems,
given by the Legendre transformation, needs  a more careful study and makes
  use of finer tools
in the case of singular Lagrangians. In this case, constraint functions
determining the submanifold in which the dynamical equation has a consistent
solution will appear. Moreover, in the Lagrange approach there will be more
constraints functions determining the submanifold in which the
dynamics admits a
solution that is the restriction of a second order differential equation vector
field. It has been  shown that the relation among  the constraint functions
arising  in the Lagrangian formalism and those of the Hamiltonian one can be
established by means of a differential operator $K$, first introduced
by Kamimura \cite{\kamimura} and later used by
Batlle {\sl et al\/} \cite{\batlleetal}, and whose geometric
interpretation was given in \cite{\GraciaPons}
and \cite{\CarinenaLopez}. For a recent review of these objects see
\cite{\GraciaPonsdos}. The theory of sections along maps
is the key point for establishing the operator $K$. In fact, vector
fields along
a map, or relative vector fields along a map according to \cite{\PugliVino}, simplify and clarify most constructions in classical mechanics
\cite{\CarinenaFer, \CarinenaLoMari,\CarinenaLoMarii} and they have
recently been
used in classical field theories \cite{\catalanes}.

On the other hand, the necessity of incorporating anticommuting
variables for describing
dynamical
systems with fermionic degrees of freedom has lead to the development
of the so-called
supermechanics
\cite{\IbortMarin}. Moreover, it has been shown to be quite useful not only in
physics but also in mathematics, particularly
in the study of the geometry associated to a Lie algebroid, mainly
due to the Vaintrob
Theorem \cite{\Vaintrob}.

Our aim in this paper is to discuss the generalization in the graded
context of the
operator $K$, also called relative Hamiltonian vector field in \cite{\PugliVino}, 
which allows us  to relate in this way constraint functions  arising
in Hamilton formalism
  for singular systems with those of the
Lagrange approach. Our intention here is not to do a complete description
  of the theory of constraints in supermechanics, but rather to
introduce some elements
to convince the reader that this theory may be developed along
parallel lines to the theory
of constraints in classical mechanics. The main difficulty in this
  enterprise lies in that
the information of a graded manifold is encoded in the sheaf of superfunctions,
instead of the underlying manifold.
Indeed, in the transition to the supermechanics setting the use of the
concepts of sections along a map
is even more necessary because of the inconvenience of working with
points in graded geometry.
Thus one is forced to take an algebraic approach, which replaces all
the intrinsic constructions
that are based on points of the manifold in the classical case. The interesting
point is that this is accomplished  by using vector
fields and forms along a morphism of supermanifolds in the same way they
where used in \cite{\CarinenaLoMari,\CarinenaLoMarii}\ in the
classical setting.
See \cite{\Hector-\Hectorv}\ for details.

    The paper is
  organized as follows:
Section 2 is devoted  to set  our
notation and, for the reader convenience, we describe the material from
the theory of graded manifolds that will be used in later Sections.
In particular, we recall the concepts  of vector fields
and graded forms along a morphism, and particular examples are given.

In Section 3 we introduce, in the graded context, the time evolution
operator $K$ associated to a super-Lagrangian function $L$,
and we discuss its main properties,
in order to study the Lagrangian constraints associated to a
singular Lagrangian and the connection with their Hamiltonian counterpart.
Finally, Section 4 analyzes the
projectability of Lagrangian constraint functions using the operator $K$.

\beginsection  2. Basic notation and background.

\medskip

Naturally, the arena to develop Lagrangian or Hamiltonian
supermechanics will be a suitable generalization, to the
graded context, of the tangent and cotangent manifold of the
configuration space. Surprisingly enough, even this requires
some attention. The point is that the superobjects that have
the right geometrical structure: the tangent or cotangent
superbundles, introduced by S\'anchez--Valenzuela in
\cite{\Sanchez}, are too big, as their dimensions are
$(2m+n,2n+m)$, if the dimension of the starting graded manifold
$\M = (M,\A)$ (the configuration superspace) is $(m,n)$. This
can be fixed by considering the subsupermanifolds of dimension
$(2m,2n)$, introduced  by Ibort and Mar\1n in
\cite{\IbortMarin}, which, nonetheless, do not have all the
geometrical richness that one is used to; for instance,
supervector fields, that is, derivations of $\A$, can be
considered as section of the tangent superbundle $ST\M$, but not
of the tangent supermanifold $T\M$. Thus, it is advisable to
define and study the main properties of the relevant objects in
$ST\M$, but to perform the computations and the interpretations
after the restriction to $T\M$~\cite{\Hectoriii}.

   For the reader convenience and to fix the notation, we shall
describe the main objects that give the geometry of the tangent
and cotangent bundles in the graded context, and refer the reader
to~\cite{\Hectoriii}\ for details.  Through out we shall be
working with supermanifolds in the sense of Kostant
\cite{\Kostant}\ and Le\u{\i}tes \cite{\Leites}.

A {\it supervector bundle} is a quadruplet
$\set{(E,\A_E),\Pi,(M,\A_M),V_S}$ such that $V$ is a real
$(r,s)$--dimensional supervector space, $\Pi \: (E,\A_E)
\to(M,\A_M)$ is a  submersion of graded  manifolds,
and every $q\in M$ lies in a coordinate neighbourhood
$\U \subseteq M$ for which an isomorphism $\Psi_\U$ exists
making the following diagram commutative:
$$
\matrix{
\Bigr(\pi^{-1}(\U),\A_E\bigr(\pi^{-1}(\U)\bigl)\Bigl)
&\harr{\displaystyle \Psi_\U}{}
&\bigr(\U,\A_M(\U)\bigl) \x V_S    \cr
\varr{\displaystyle\Pi}{}
&& \varr{}{\displaystyle P_1}  \cr
\bigr(\U,\A_M(\U)\bigl)
&\harrt{}{}  &\bigr(\U,\A_M(\U)\bigl)\qquad .\cr}
$$
Here $V_S := S(V\oplus {\overline \Pi} V)$ where ${\overline \Pi}$ is 
the change of
parity  functor~\cite{\Leites,\Manin}, hence
$({\overline \Pi} V) = ({\overline \Pi} V)_0 \oplus ({\overline \Pi} 
V)_1$, where $({\overline \Pi} V)_i =
V_{i+1}$ for $i = 0,1$, and $S(V)$ is the affine supermanifold
$$
S(V) := \bigl
(V_0,C^\infty(V_0) \ox \bw(V_1^*)\bigr)\ .
$$
Equivalence classes of supervector bundles so defined are in a
one--to--one correspondence with equivalence classes of locally
free sheaves of $\A_M$--modules over $\M$ of rank $(r,s)$. The
tangent and cotangent superbundles are the superbundles
corresponding to the sheaves $\X(\A) :=\Der \A$ and
$\Om^1(\A) := \X(\A)^*$ respectively.

The main reason for considering the tangent superbundle
$\{(STM,ST\A),\Tau,(M,\A)\}$, and supervector bundles in
general~\cite{\Sanchez}, is that their geometrical sections
are in a one--to--one correspondence with the sections of the
corresponding locally free sheaf of graded $\A$--modules;
in our case, with the sections of the sheaf $\Der \A$, in other
words, with the supervector fields over $\M$.  Unfortunately,
the use of the parity functor ${\overline \Pi}$ introduces some unwanted
supercoordinates; the elimination of these coordinates lead
to the tangent and cotangent supermanifolds~\cite{\Hectoriii}.

Supervector fields, or graded forms, along a morphism are our main
tool to describe supermechanics, in fact all the relevant objects
can be defined as such \cite{\Hector--\Hectorv}.  This is so
because of their algebraic nature and because the information of
a graded manifold is concentrated in the algebraic part, that is
in the sheaf of superalgebras.

If $\Phi =(\phi,\phi^*)\: (N,\B) \to (M,\A)$ is a morphism of
graded manifolds, a homogeneous {\it supervector field along}
$\Phi$ is a morphism of sheaves over $M$, $X\: \A \to \Phi_*\B$
such that for each open subset $\U$ of $M$
$$
X(fg) =
X(f) \, \phi^*_\U(g) + \srule{X}{f} \phi^*_\U(f) \, X(g)\ ,
$$
whenever $f\in \A(\U)$ is homogeneous of degree $\abs{f}$. The
sheaf of supervector fields along $\Phi$ will be denoted by
$\X(\Phi)$. If $X$ is a supervector field on $(M,\A)$, an
example of an element in $\X(\Phi)$ is given by
$$
\hat X := \phi^* \circ X \ .
\eqlabel\uno
$$

Similarly if $Y\in \X(\B)$, then
$$
T\phi(Y) := Y \circ \phi^*,
\eqlabel\dos
$$
also belongs to $\X(\Phi)$. We say that $Y$ is {\it projectable
with respect to} $\Phi$ if there exists $X\in\X(\A)$ such that
$T\phi(Y) = \hat X$. Sometimes, we also say that $X$ and $Y$ are
$\Phi$-{\it related}.

   $\X(\Phi)$ is a locally free sheaf of $\Phi_*\B$--modules over
$M$ of rank $(m,n) = \dim\M$; a local basis of $\X(\Phi)(\U)$ is
given by
$$
\pd{\hat q^i} := \widehat\pd{q^i}\,,
\qquad\qquad
\pd{\hat\th^\a} := \widehat\pd{\th^\a}\ ,
$$
if $(q^i,\th^\a)$ ($1\le i\le m$, $1\le \a\le n$), are local
supercoordinates on $\U\subset\M$~\cite{\Hector}.

The sheaf of {\it graded 1--forms along} $\Phi$ is the sheaf of
$\phi_*\B$--modules
$\Om^1(\Phi) := \X(\Phi)^* = \Hom(\X(\Phi),\phi_*\B)$. If $\om$ is
a graded 1--form on $\M$, $\hat\om$ defined by
$$
\hat\om(\hat X) := \phi^* \circ \om (X)\,,
\qquad \forall X\in\X(\A_M)\ ,
\eqlabel\tres
$$
belongs to $\Om^1(\Phi)$.  Moreover, a local basis is given by
the elements $d\hat q^i := \widehat{dq^i}$ and $d\hat\th^\a :=
\widehat{d\th^\a}$.  On the other hand, if $\om$ is a graded
1--form along $\Phi$, then $\phi^\sharp\om$ given by
$$
\phi^\sharp\om (Y):= \om \bigl(T\phi(Y)\bigr)\,,
\qquad\forall Y\in\X(\B)\ ,
\eqlabel\cuatro
$$
is a graded 1--form on $\ENE$.  As a matter of fact, it is
possible to classify the graded 1--forms on $\ENE$ that come from
graded 1--forms along $\Phi$, when $\Phi$ is a submersion. The
result is that $\Om^1(\Phi)$ is isomorphic to the
$\phi_*\B$--modulo of $\Phi$--semibasic 1--forms on
$\ENE$~\cite{\Hector}. Naturally, we can extend \eq\tres\ and
\eq\cuatro\ to arbitrary graded forms.

   Let $\Ec = \set{(E,\A_E),\Pi,(M,\A_M),V_S}$ be a
superbundle. A local {\it section of $\cal{E}$ along} $\Phi$, over
an open subset $\U$ of $M$, is a morphism,
$\Si =
(\si,\si^*)\:\Big(\phi^{-1}(\U),\B\bigl(\phi^{-1}(\U)\bigr)\Bigr)
\to  \Big(\pi^{-1}(\U),\A_E\bigl(\pi^{-1}(\U)\bigr)\Bigr)$,
satisfying the condition $\Phi_\U = \Pi_\U \circ \Si_\U$,
where the subscript $\U$ means the restriction of the morphism to
the corresponding open graded submanifold.  The set of such
sections is denoted by $\Ga_\Phi(\Pi|_\U)$. When $(E,\A_E)$ is
the tangent or the cotangent superbundle these sections are
in a one--to--one correspondence with supervector fields and
graded 1--forms along $\Phi$, respectively~\cite{\Hectoriii}.

In the case when the morphism $\Phi$ coincides with the
projection $\Pi$ of the supervector bundle, the identity morphism
on $\cal{E}$ gives a canonical section. In the tangent superbundle
$\{ST\M,\Tau,\M\}$ the supervector field along $\Tau = (\tau,
\tau^*) $ that corresponds to the canonical section is called the
{\it total time derivative operator} and is denoted by
$\T$. Whereas the $\Pi$--semibasic graded 1--form on $ST^*\M$
associated to the graded 1--form along $\Pi$, corresponding to
the canonical section of the cotangent superbundle
$\{ST^*\M,\Pi= (\pi, \pi^*),\M\}$, is called {\it the canonical
Liouville 1--form} on $ST^*\M$ and will be denoted by $\Th_0$.
The restrictions of $\T$ and $\Th_0$ to the tangent and cotangent
supermanifolds, that will be denoted in the say way, can be
written, in the natural supercoordinates of these
supermanifolds~\cite{\Hectoriii}\ associated to the
supercoordinates $q^i,\th^\a$ of $\M$ on $\U \subset M$, as
$$
\T = \sum_{i=1}^m v^i\pd{\hat q^i}
+\sum_{\a=1}^n \z^\a\pd{\hat\th^\a},
\sepword{and}\quad
\Th_0 = \sum_{i=1}^m p^i\, dq^i +
\sum_{\a=1}^n \eta^\a \,d\th^\a.
\eqlabel\cinco
$$
The reason to consider these restrictions is that although
$\Th_0$ is formally equal to the canonical 1--form of the
cotangent bundle in non--graded geometry, it turns out that
the graded 2--form $-d\Th_0$ is always degenerate, whereas the
restriction of $\Th_0$ to the cotangent supermanifold $T^*\M$
gives a non--degenerate graded 2--form $\om_0 := -d\Th_0$.

Using the abbreviation $T\AU$ for $T\A\bigl(\tau^{-1}(U)\bigr)$, we
associate to each superfunction $f\in\AU$ the superfunction
$f^V\in T\AU$ defined by
$$
f^V := \sum^m_{i=1} \pdf{F}{q^i}v^i
+ \sum^n_{\a=1} \pdf{F}{\th^\a}\z^\a,
\eqlabel\seis
$$
where $F:= \tau^*(f)\in T\AU$.  It turns out that a supervector
field $Y$ on $T\M$ is determined by its action on the
superfunctions $f^V$. Thus, if $X$ is a supervector field on $\M$,
or a supervector field along $\Tau$, its  {\it vertical lift} is
the supervector field $X^V$ on $T\M$ defined by
$$
X^V(f^V) = \tau^*\bigr(X(f)\bigl)\,,
\qquad \forall f\in\A\ .
$$
In local supercoordinates, if
$X = \sum_{i=1}^m X^i\pd{\hat q^i}
+ \sum_{\a=1}^n \chi^\a\pd{\hat\th^\a}$, then
$$
X^V = \sum_{i=1}^m X^i\pd{v^i}
+ \sum_{\a=1}^n \chi^\a\pd{\z^\a}.
$$
We are now in a position to introduce the superobjects
corresponding  to the objects that determine the geometry of the
tangent manifold~\cite{\CrampinPirani}: the {\it vertical
superendomorphism} is the graded tensor field of type $(1,1)$
$S\: \X(T\A) \to \X(T\A)$  defined by
$$
S(Y) := \bigl(T\tau(Y)\bigr)^V.
\eqlabel\siete
$$
On the other hand, the {\it Liouville supervector field} $\De$
is the vertical lift of the total time derivative:
$$
\De := \T^V.
$$
If $Y = \sum_{i=1}^m Y^i\pd{q^i}
+ \sum_{i=1}^m{\cal Y}^i\pd{v^i}
+ \sum_{\a=1}^n \Upsilon^\a\pd{\th^\a}
+ \sum_{\a=1}^n \Xi^\a\pd{\z^\a}$ then,
$$
S(Y) = \sum_{i=1}^m Y^i\pd{v^i}
+ \sum_{\a=1}^n \Upsilon^\a\pd{\z^\a}.
$$
In analogy with ordinary Lagrangian mechanics, the {\it graded
Cartan 1 and 2--forms} associated to a given Lagrangian
superfunction $L$ in $T\A$ are defined by
$$
\Th_L := dL \circ S \sepword{and}\quad \om_L := -d\Th_L.
$$
Since $\Th_L$ is a $\Tau$--semibasic graded 1--form, and $\Tau$
is a submersion, it has associated a unique graded 1--form
$\widehat{\Th}_L$ along $\Tau$. In analogy with non--graded
geometry, see~\cite{\CarinenaLoMarii}, the restriction to $T\M$
of the  section $FL \: ST\M \to ST^*\M$ along $\Tau$ that
corresponds to the graded 1--form $\widehat{\Th}_L$ is called the
{\it super--Legendre transformation}. If $L$ is even, locally,
$FL =  (fl,fl^*)$ is determined by the morphism of superalgebras
$fl^*\: T^*\AU \to T\AU$ described by the relations:
$$
\leqalignno{
q^i \mapsto q^i\,,  \qquad\qquad  & \th^\a \mapsto \th^\a\ ,
& (\ecnum) \cr
p^i \mapsto \pdf{L}{v^i}\,,  \qquad\quad
& \eta^\a \mapsto -\pdf{L}{\z^\a}\ .    \cr}
\eqlabel\ocho
$$
For more details on the super--Legendre transformation
see~\cite{\Hectoriii}.

\beginsection  3. The time evolution operator.

\medskip

In Lagrangian supermechanics the dynamics of a system
$(T\M, \om_L, L)$, associated to a regular Lagrangian $L\in T\M$,
is given by a vector field $\Ga \in \X(T\M)$ satisfying the {\it
dynamical equation}
$$
i_\Ga \om_L= d\,E_L\ .
\eqlabel\one
$$
This uniquely defined vector field $\Ga$ satisfies, automatically, the
second order condition \cite{\IbortMarin}, which can be stated in
several equivalent ways. A very convenient one, suitable to
generalization to higher orders \cite{\Hector,\Hectorv}, is
$$
\Ga \circ \tau^* = \T\ .
\eqlabel\two
$$
To abbreviate, we say that $\Ga$ is a SODE vector field (Second
Order Differential Equation). When the super-Lagrangian $L$ is
singular both the existence and  uniqueness of $\Ga$ are in
jeopardy, and it is necessary to consider a submanifold of $T\M$
where \eq\one\ holds.  Moreover, even on this submanifold \eq\two\
may fail, so both conditions have to be considered separately.
Motivated by these issues we consider the following definition:

\declare Definition 3.1.
The {\bf time evolution operator} $K \: T^*\A \to T\A$, associated
to a Lagrangian super-function, $L\in T\M$, is the unique supervector field
along the super-Legendre transformation $FL$ satisfying the {\it
dynamical condition}
$$
i_K \om_0 = d\,E_L\ ,
\eqlabel\three
$$
and the {\it second order condition}
$$
K \circ \pi^* = \T\ .
\eqlabel\four
$$
Since $\T$ is even, \eq\four\ implies that $K$ is also even, hence
$i_K \: \Om(T^*\A) \to \Om(T\A)$ is the unique $FL^*$--derivation
of bidegree $(-1,0)$~\cite{\Hector} defined by
$$
i_K f = 0  \sepword{and}\quad  i_K df = K(f)\ .
$$
In particular, one has
$$
i_K( \om \w \mu) =  i_K \om \w FL^*\mu
+ (-1)^{\abs{\om}}  FL^*\om \w i_K \mu\ ,
\eqlabel\five
$$
when $\om$ is homogeneous.  Conditions \eq\three\ and \eq\four\
are the same conditions as those used in \cite{\GraciaPons}
to define, in the non graded context, the time evolution
operator, written in the algebraic language of operators to avoid
the use of points of the underlying manifold.

\proclaim Proposition 3.1.
There exist a unique supervector field along $FL$, $K\in \X(FL)$ satisfying
\eq\three\ and \eq\four.

\Proof
In the local supercoordinates on $T\AU$ and $T^*\AU$ naturally
associated to those on $\AU$, see~\cite{\Hectoriii}, equation
\eq\four\ implies
$$
K(q^i) = K \circ \pi^*(q^i) = \T(q^i) = v^i
\sepword{and}\quad
K(\th^\a) = K \circ \pi^*(\th^\a) = \T(\th^\a) = \eta^\a.
\eqlabel\six
$$
Assume now that $\abs{L} = 0$. Since $\om_0 = \sum_i dq^i \w dp^i
- \sum_\a d\eta^\a \w d\th^\a$, then, using \eq\five\ and
\eq\ocho,
$$
\leqalignno{
i_K \om_0 &=
\sum_i K(q^i) \, d\bigl(fl^*(p^i)\bigr)
- \sum_i K(p^i) \, d\bigl(fl^*(q^i)\bigr)  \cr
&\quad - \sum_\a K(\th^\a) \, d\bigl(fl^*(\eta^\a)\bigr)
- \sum_\a K(\eta^\a) \, d\bigl(fl^*(\th^\a)\bigr)  \cr
&= \sum_j\biggl(\sum_i v^i \sopd{L}{q^j}{v^i}
+ \sum_\a \zeta^\a \sopd{L}{q^j}{\th^\a}
- \sum_i \de_{ij}\, K(p^i) \biggr)\, dq^j  \cr
&\quad + \sum_j\biggl( \sum_i v^i \sopd{L}{v^j}{v^i}
+ \sum_\a \zeta^\a \sopd{L}{v^j}{\zeta^\a} \biggr)\, dv^j
& (\ecnum) \cr
&\quad - \sum_\b\biggl(\sum_i v^i \sopd{L}{\th^\b}{v^i}
- \sum_\a \zeta^\a \sopd{L}{\th^\b}{\zeta^\a}
+ \sum_\a \de_{\a\b}\, K(\eta^\a) \biggr)\, d\th^\b  \cr
&\quad - \sum_\b\biggl( \sum_i v^i \sopd{L}{\zeta^\b}{v^i}
- \sum_\a \zeta^\a \sopd{L}{\zeta^\b}{\zeta^\a} \biggr)\,
d\zeta^\b. \cr}
\eqlabel\seven
$$
On the other hand, as $\De = \T^V$, \eq\cinco\ implies $\De(L) =
\sum_i v^i \pdf{L}{v^i} + \sum_\a \zeta^\a \pdf{L}{\zeta^\a}$,
therefore
$$
\leqalignno{
dE_L &=
\sum_j\biggl(\sum_i v^i \sopd{L}{q^j}{v^i}
+ \sum_\a \zeta^\a \sopd{L}{q^j}{\th^\a}
- \sum_i\de_{ij}\, \pdf{L}{q^i} \biggr)\, dq^j  \cr
&\quad + \sum_j\biggl( \sum_i v^i \sopd{L}{v^j}{v^i}
+ \sum_\a \zeta^\a \sopd{L}{v^j}{\zeta^\a} \biggr)\, dv^j
& (\ecnum) \cr
&\quad - \sum_\b\biggl(\sum_i v^i \sopd{L}{\th^\b}{v^i}
- \sum_\a \zeta^\a \sopd{L}{\th^\b}{\zeta^\a}
- \sum_\a \de_{\a\b}\, \pdf{L}{\th^\a} \biggr)\, d\th^\b  \cr
&\quad - \sum_\b\biggl( \sum_i v^i \sopd{L}{\zeta^\b}{v^i}
- \sum_\a \zeta^\a \sopd{L}{\zeta^\b}{\zeta^\a} \biggr)\,
d\zeta^\b. \cr}
\eqlabel\eight
$$
Thus, if \eq\three\ holds \eq\seven\ and \eq\eight\ give
$$
K(p^i) = \pdf{L}{q^i}
\sepword{and}\quad
K(\th^\a) = - \pdf{L}{\th^\a}\ .
\eqlabel\nine
$$
This together with \eq\six\ imply the uniqueness of $K$.
Moreover, it is easy to verify that \eq\six\ and \eq\nine\ do
define a supervector field along $FL$, which proves the
proposition  when $L$ is even; the odd case is proved in the
same way but \eq\nine\ are different.   \qed

In the non graded case the time evolution operator was defined
in \cite{\CarinenaLopez}\ using the generalized Hamiltonian
system defined on the mixed space $TM \o+ T^*M$
\cite{\Skinner,\SkinnerRusk} as follows: given a Lagrangian
function $L\in TM$ we consider on $TM \o+ T^*M$ the 2-form
$\Om := \pr_2^* \om_0$ and the function $D:= \<\pr_1,\pr_2> -
\pr_1^*L$, where $\pr_i$ denotes the projection of $TM \o+ T^*M$
onto the $i$--th factor. If $W$ denotes the graph of the Legendre
transformation
$$
W= \set{ (v,p) \in TM \o+ T^*M : p = FL(v)}\ ,
$$
then the map $\Bar{FL} \: TM     \to W$ given by $\Bar{FL}(v) =
\bigl(v , FL(v) \bigr)$ is a diffeomorphism whose inverse is
$\pr_1|_W$ \cite{\Skinner,\SkinnerRusk}. To simplify the notation
we shall also denote the restriction of $\pr_i$ to $W$ by $\pr_i$.
The time evolution operator $\widetilde K \: C^\infty(T^*M) \to
C^\infty(TM)$ is defined by
$$\widetilde K := \Bar{FL}^* \circ Z
\circ \pr_2^*\ \eqlabel\defKtilde$$ where $Z$ is any vector field on 
$W$ satisfying
$$
i_Z\, \Om = dD.
\eqlabel\ten
$$
Now, to prove that both definitions agree, we first  notice that
$$
\Bar{FL}^*D(v) = D\bigl(v , FL(v) \bigr)
= \<v, FL(v)> - \Bar{FL}^* \circ \pr_1^*  L(v)
= \De L(v) - L(v) = E_L(v)\ ,
\eqlabel\eleven
$$
and
$$
\Bar{FL}^* \Om = \Bar{FL}^* \circ \pr_2^* \om_0
= (\pr_2 \circ \Bar{FL})^* \om_0 = FL^* \om_0 = \om_L\ .
\eqlabel\twelve
$$

\proclaim Proposition 3.2.
If $X$ is a vector field on $TM$ such that $\pr_1^* \circ X =
Z \circ \pr_1^*$, then $i_X \om_L = dE_L$ and $X \circ \tau^* =
\T$.

\Proof
Since $ X \circ \Bar{FL}^* = \Bar{FL}^* \circ Z$, equations
\eq\eleven\ and \eq\twelve\ yield
$$
i_X \om_L = i_X \circ \Bar{FL}^* \Om = \Bar{FL}^* \circ i_Z \Om
= \Bar{FL}^*(dD) = dE_L.
$$
To prove the second assertion, for each $\a \in \Om^1(M)$ let
$Y_\a \in \X(T^*M)$ be the unique vector field such that
$$
i_{Y_\a} \om_0 = \pi^* \a\ ,
$$
and let $Z_\a \in \X(W)$ be such that $Z_\a \circ \pr_1^* = 0$ and
$Z_\a \circ \pr_2^* = \pr_2^* \circ Y_\a$, then
$$
i_{Z_\a} \Om = i_{Z_\a} \circ \pr_2^* \om_0
= \pr_2^* \circ i_{Y_\a} \om_0 = \pr_2^* \circ \pi^* \a
= (\tau \circ \pr_1)^* \a\ ,
$$
therefore, using \eq\tres
$$
\leqalignno{
\Om(Z , Z_\a) &= -\Om(Z_\a , Z) = - i_{Z_\a} \Om Z
= - (\tau \circ \pr_1)^* \a(Z)  & (\ecnum) \cr
&= - \hat\a(Z \circ \pr_1^* \circ \tau^*)
= - \hat\a(\pr_1^*\circ X \circ \tau^*)\ .  \cr}
$$
On the other hand, since $Z_\a \circ \pr_1^* = 0$,
$$
\Om(Z , Z_\a) = i_{Z_\a} i_Z \Om = i_{Z_\a} dD
= i_{Z_\a} d\<\pr_1,\pr_2> - Z_\a \circ \pr_1^*L
= i_{Z_\a} d\<\pr_1,\pr_2>\ .
$$
Now, if $q^i$ are local coordinates on $M$, $v^i$ and $p^i$ are
the corresponding local coordinates on $TM$ and $T^*M$
respectively, and $\a = \sum_i \a_i dq^i$, then a simple
computation in local coordinates, using that $Z_\a \circ \pr_2^* =
\pr_2^* \circ Y_\a$ and \eq\tres, gives
$$
i_{Z_\a} d\<\pr_1,\pr_2>
= - \sum_i \pr_1^* v^i \, \pr_1^* \circ \tau^* \a_i
= - \hat \a(\pr_1^* \circ \T)\ .
$$
Thus, $\hat\a(\pr_1^*\circ X \circ \tau^*) = \hat \a(\pr_1^* \circ
\T)$. Since $\a$ is arbitrary and $\pr_1^*$ is injective, it
follows that $X \circ \tau^* = \T$.    \qed

\proclaim Proposition 3.3. Let $L$ be a Lagrangian superfunction. Then
  $\widetilde K$ defined by  \eq\defKtilde\ coincides with $K$ as
given by  \eq\three\ and
\eq\four, i.e.
$K = \widetilde K$. This also proves that $\widetilde K$ is
independent of the choice of $Z$ satisfying \eq\ten.

\Proof
Clearly $\widetilde K$ is a vector field along $FL$, therefore it
remains to prove that $\widetilde K$ satisfies \eq\three\ and
\eq\four. By Lemma 2.1 of \cite{\Hector} and \eq\eleven
$$
i_{\widetilde K} \om_0 = i_{\Bar{FL}^* \circ Z \circ \pr_2^*} \om_0
= i_{\Bar{FL}^* \circ Z} \pr_2^*\om_0
= i_{\Bar{FL}^* \circ Z} \Om = \Bar{FL}^* \circ i_Z \Om
= \Bar{FL}^* dD = dE_L\ .
$$
On the other hand, if $X \in \X(TM)$ is such that $\pr_1^* \circ X
= Z \circ \pr_1^*$, then by Proposition 3.2
$$
\widetilde K \circ \pi^* = \Bar{FL}^* \circ Z \circ \pr_2^* \circ \pi^*
= \Bar{FL}^* \circ Z \circ \pr_1^* \circ \tau^*
=  \Bar{FL}^* \circ \pr_1^* \circ X \circ \tau^*
= X \circ \tau^* = \T\ .  \qed
$$

We point out that our arguments were cast so as to hold also
in the graded context. The only technical point is to define the
Whitney sum of supervector bundles, which can be done exactly as
in the classical case~\cite{\tesis}. Moreover, the properties of
$K$, discussed in what follows, will also be written in such a way
so as to hold on supermanifolds by adding the supercoordinates that
anticommute. Nevertheless, to simplify the notation, we shall work
on a differential manifold $M$.

Notice that if $X$ and $Z$ are as in Proposition 3.2, and
$g \in \Coo(TM)$ is a $FL$-projectable function, say $g =
FL^*(h)$, then
$$
X(g) = X\circ FL^*(h) = X\circ \Bar{FL}^* \circ \pr_2^* (h)
= \Bar{FL}^* \circ Z \circ \pr_2^* (h) = K(h)\ ,
$$
therefore the operator $K$ gives the time evolution for this
kind of functions, and provides a reason for the name of the
operator.

The main property of $K$ is that its action on Hamiltonian
constraints generates the Lagrangian constraints
\cite{\batlleetal,\CarinenaLopez}.  Before we see how this goes, we shall
introduce another operator that is also used to compare
Hamiltonian and Lagrangian constraints \cite{\CarinenaLoRo}, but
again we define it using an algebraic approach that can be
generalized to the graded context. Since  a vector field on $TM$
is determined by its action on the maps $f^V$ defined in
\eq\seis, we associate to each $U \in \X(FL)$ the vector field on
$TM$ defined by
$$
\widetilde R_L U(f^V) := U \circ \pi^*(f) \qquad \forall f
\in\Coo(M).
$$
Now, if $Y\in\X(T^*M)$, then $FL^*\circ Y \in \X(FL)$, so we can
define an  operator $R_L \: \X(T^*M) \to \X(TM)$ by
$$
R_L(Y) := \widetilde R_L (FL^*\circ Y)\ .
$$
When $Y = \sum_{i=1}^m (Y^i \pd{q^i} + \Y^i
\pd{p^i})$, then $R_L(Y) = \sum_{i=1}^m Y^i
\pd{v^i}$.  In particular, if $h\in \Coo(T^*M)$, and
$Y_h$ is the vector field such that
$$
i_{Y_h}\om_0 = dh\ ,
\eqlabel\thirteen
$$
then
$$
R_L(Y_h)
= \sum_{i=1}^m FL^*\Bigl(\pdf{h}{p^i}\Bigr) \pdf{}{v^i}\ .
\eqlabel\fourteen
$$
Moreover, if $X\in\X(TM)$ is a vector field such that
$X \circ \tau^* = FL^* \circ Y \circ \pi^*$, then by \eq\siete
$$
R_L( Y)(f^V)  = X\circ \tau^*(f) = (T\tau X)^V(f^V) = S(X)(f^V)\ ,
$$
so $R_L (Y )= S(X)$.

\proclaim Lemma 3.4.
For each $h \in \Coo(T^*M)$ there are vector fields $X_h$ on
$TM$ such that $S(X_h) = R_L(Y_h)$, where $Y_h$ is the vector
field defined by \eq\thirteen.

\Proof
First we choose $Z_h \in \X(W)$ such that $Z_h \circ \pr_2^* =
\pr_2^* \circ Y_h$, then we take $X_h$ such that
$Z_h \circ \pr_1^* = \pr_1^* \circ X_h$. Now,
$$
\leqalignno{
\pr_1^* \circ X_h \circ \tau^* &= Z_h \circ \pr_1^*
\circ \tau^* = Z_h \circ \pr_2^* \circ \pi^*
& (\ecnum) \cr
&= \pr_2^* \circ Y_h \circ \pi^*
= \pr_1^* \circ FL^* \circ  Y_h \circ \pi^*. \cr}
$$
Since $\pr_1^*$ is injective $X_h \circ \tau^* =
FL^* \circ  Y_h \circ \pi^*$, so by the comment before the
statement $S(X_h) = R_L(Y_h)$.  \qed

Note that $X_h$ is by no means unique, but clearly the difference
of two such vector fields is $\tau$-vertical.

\proclaim Proposition 3.5.
For each $h \in \Coo(T^*M)$
$$
K(h) = i_{X_h} [i_\Ga \om_L - dE_L] + i_\Ga d(FL^* h)\ ,
\eqlabel\fifteen
$$
where $X_h$ is any vector field such that $S(X_h) = R_L(Y_h)$,
and $\Ga \in \X(TM)$ is an arbitrary SODE vector field.

\Proof
Given $X_h$ and $\Ga$, we choose vector fields  $U_h$ and $V$ in
$\X(W)$ such that $U_h \circ \pr_1^* = \pr_1^* \circ X_h$, and
$V \circ \pr_1^* = \pr_1^* \circ \Ga$. If $Z_h$ is as in the proof
of the previous lemma, then,
$$
i_{Z_h} \Om = i_{Z_h} \pr_2^*\om_0
= \pr_2^* \circ i_{Y_h} \om_0 = \pr_2^*(dh)\ ,
\eqlabel\sixteen
$$
so
$$
\leqalignno{
K(h) &= \Bar{FL}^* \circ Z \circ \pr_2^* (h)
= \Bar{FL}^* \circ i_Z \circ i_{Z_h} \Om
= - \Bar{FL}^*  \circ i_{Z_h} \circ i_Z \Om   \cr
&= - \Bar{FL}^*  \circ i_{Z_h - U_h} \circ i_{Z - V} \Om
- \Bar{FL}^*  \circ i_{Z_h} \circ i_V \Om
& (\ecnum) \cr
&\qquad  + \Bar{FL}^*  \circ i_{U_h} \circ i_V \Om
- \Bar{FL}^*  \circ i_{U_h} \circ i_Z \Om\ .  \cr}
\eqlabel\seventeen
$$
By the proof of Lemma 3.4, $Z_h - U_h$ is $p$-vertical, where
$p\: W \to M$ is the canonical projection. But   $Z - V$ is also
$p$-vertical since $Z(q^i) = v^i = V(q^i)$, hence
$i_{Z_h - U_h} \circ i_{Z - V} \Om = 0$.

On the other hand, by \eq\eleven\ and \eq\twelve,
$$
\leqalignno{
\Bar{FL}^*  \circ i_{U_h} \circ i_Z \Om
&= i_{X_h} \circ \Bar{FL}^* \circ i_Z \Om
= i_{X_h} \circ \Bar{FL}^* (dD) = i_{X_h} dE_L\ ,
& (\ecnum) \cr
\Bar{FL}^*  \circ i_{U_h} \circ i_V \Om
&= i_{X_h} \circ \Bar{FL}^* \circ i_V \Om
= i_{X_h} \circ i_\Ga \Bar{FL}^*\Om
= i_{X_h} \circ i_\Ga \om_L\ .  \cr}
\eqlabel\eighteen
$$
Finally, using \eq\sixteen,
$$
\leqalignno{
\Bar{FL}^*  \circ i_{Z_h} \circ i_V \Om
&= -\Bar{FL}^* \circ i_V  \circ i_{Z_h} \Om
= - i_\Ga \circ \Bar{FL}^* \circ i_{Z_h} \Om
& (\ecnum) \cr
&= - i_\Ga \circ \Bar{FL}^*(d\pr_2^*h)
= - i_\Ga d(\pr_2 \circ \Bar{FL})^* h
= - i_\Ga d(FL^* h)\ .
\cr}
\eqlabel\nineteen
$$
Plugging \eq\eighteen\ and \eq\nineteen\ into \eq\seventeen\ we
obtain \eq\fifteen.  \qed

When $h$ is a Hamiltonian constraint, so $FL^* h = 0$,
\eq\fifteen\ reduces to
$$
K(h) = i_{X_h} [i_\Ga \om_L - dE_L]\ ,
$$
and the right hand side is a Lagrangian constraint
\cite{\CarinenaLoRo,\MunozRoman}; actually, it is the constraint
associated to $h$ through the operator $R_L$. Thus, the operator
$K$ reproduces the Lagrangian constraints, while the operator
$R_L$ provides the non arbitrary part of the vector fields
associated to a given constraint.

\bigskip

\beginsection 4. Projectability of constraints

\medskip

In order to analyze the projectability of these constraints we consider
the following lemmata.

\proclaim Lemma 4.1.
The energy function $E_L$ is $FL$-projectable.

\Proof
We have to proof that $E_L$ is annihilated by all the elements of
$\ker FL_* = \set{X\in \X(TM) \: X \circ FL^* = 0}$. Since $\tau^*
= FL^* \circ \pi^*$, it is clear that vector fields  in $\ker
FL_*$ are $\tau$-vertical. In particular, $i_X \th_L = 0$ when $X
\in \ker FL_*$, as $\th_L = dL \circ S$. On the other hand, if
$\Ga$ is a SODE vector field, and $X \in \ker FL_*$, then
$[X,\Ga](q^i) = X \Ga (q^i)= X(v^i)$, so
$$
S([X,\Ga]) = X\ .
\eqlabel\twenty
$$
Thus, if $X \in \ker FL_*$,
$$
X(E_L) = \L_X(i_\Ga \th_L -L) =
i_{[X,\Ga]} \th_L -i_\Ga \L_X \th_L - X(L)
= dL \circ S([X,\Ga]) - X(L) = 0\ .  \qed
$$

Consider the set $\Mg = \set{X\in \X(TM) \: S(X) \in \Vker \om_L}$, where
$\Vker \om_L$ denotes the set of those vector fields in
$\ker \om_L$ that are $\tau$-vertical, that is $\Vker \om_L :=
\X^v(TM) \cap \ker \om_L$.

\proclaim Lemma 4.2.
$\Mg$ coincides with
  the orthogonal complement of
the set of vertical vector fields with respect to $\om_L$,
$\Mg = \bigl(\X^v(TM)\bigr)^\perp$.

\Proof
Since
$$
\om_L(X, S(U)) = - \om_L(S(X), U)
\eqlabel\tone
$$
for arbitrary $X$ and $U$ in $\X(TM)$ \cite{\CrampinPirani,
Section 13.8}\ (or see \cite{\IbortMarin}\ for a proof in the
graded context), then
$$
\leqalignno{
\bigl(\X^v(TM)\bigr)^\perp
&= \set{X\in \X(TM) \: \om_L(X,V)=0 \,\,\hbox{for all}\,\,
V\in \X^v(TM)}
\cr &= \set{X\in \X(TM) \: \om_L(X,S(U))=0 \,\,\hbox{for all}\,\,
U\in \X(TM)}  & (\ecnum) \cr
&= \set{X\in \X(TM) \: \om_L(S(X),U)=0 \,\,\hbox{for all}\,\,
U\in \X(TM)} = \Mg\ .   \qed \cr}
$$

\proclaim Lemma 4.3.
The kernel of the Cartan 2--form associated to a Lagrangian superfunction is
$\Mg^\perp$, i.e.  $\ker \om_L = \Mg^\perp$.

\Proof
Obviously $\ker \om_L \subseteq \Mg^\perp$. On the other hand,
notice that $\om_L(V_1,V_2) =0$ if $V_1$ and $V_2$ are
$\tau$-vertical. This means that $\X^v(TM) \subset
\bigl(\X^v(TM)\bigr)^\perp =\Mg$, therefore $\Mg^\perp
\subseteq \bigl(\X^v(TM)\bigr)^\perp =\Mg$.

Now, $\pd{q^i}\in \Mg$ since $S(\pd{q^i}) = \pd{v^i} \in \X^v(TM)
\subset \Mg$, therefore if $X \in \Mg^\perp$
$$
\om_L(X,\pd{q^i}) = 0\ .
$$
Furthermore, since $\Mg^\perp \subseteq \Mg$, then $S(X)\in
\Vker \om_L$, so using \eq\tone
$$
0 = \om_L(S(X),\pd{q^i}) =\om_L(X,S(\pd{q^i})) =
\om_L(X,\pd{v^i}),
$$
and  $i$ being  arbitrary, it follows that $X \in \ker \om_L$.
\qed

The importance of the set $\Mg$ lies in the following
proposition.

\proclaim Proposition 4.4.
If $h$ is a Hamiltonian constraint and $X_h$ the
corresponding vector field constructed in Lemma 3.4, then
$X_h \in \Mg$.

\Proof
By \eq\tone\ and \eq\tres
$$
\leqalignno{
\om_L(S(X_h), X) &= - \om_L(X_h, S(X))
= -\hat \om_0(X_h \circ FL^*, S(X)\circ FL^*)
& (\ecnum) \cr
&= -\hat \om_0(FL^* \circ Y_h, S(X) \circ FL^*)\ .  \cr}
$$
But for an arbitrary $Y \in \X(T^*M)$,
$$
\hat \om_0(FL^* \circ Y_h, \hat Y)
= FL^*\bigl(\om_0(Y_h, Y)\bigr)
= FL^*\bigl(df(Y)\bigr) = \Hat{df}(\hat Y)\,;
$$
therefore $\om_L(S(X_h), X) = \Hat{df}(S(X) \circ FL^*)$.  On the
other hand, using local coordinates, it is easy to
check that $\Hat{dh}(X \circ FL^*) = X\circ FL^*(h)$ for
all $X \in \X(TM)$.  Thus, if $h$ is a Hamiltonian
constraint $FL^* h = 0$, and
$$
\om_L(S(X_h), X) = \Hat{df}(S(X) \circ FL^*)
= S(X) \circ FL^*(h) = 0,
$$
so $S(X_h) \in \ker \om_L$.  \qed

\proclaim Theorem 4.5.
Let $h\in \Coo(T^*M)$ be a Hamiltonian constraint. The associated
Lagrangian constraint $C_h := K(h) = i_{X_h}[i_\Ga \om_L - dE_L]$
is $FL$-projectable if, and only if, the vector field $X_h$
constructed in Lemma 3.4 belongs to $\ker \om_L$.

\Proof
Since $E_L$ is $FL$-projectable there exist $H \in
\Coo(T^*M)$ such that $FL^* H = E_L$. Thus if $X_h \in
\ker \om_L$
$$
\leqalignno{
C_h &= i_{X_h} dE_L = X_h \circ FL^* (H)
= X_h \circ \Bar{FL}^* \circ \pr_2^* (H)
& (\ecnum) \cr
&= \Bar{FL}^* \circ Z_h \circ \pr_2^* (H)
=  \Bar{FL}^* \circ \pr_2^* \circ Y_h (H)
= FL^* \bigl(Y_h(H)\bigr)\ .  \cr}
$$
(Here we are using the notation as in the proof of Lemma 3.4).

Conversely, assume that $C_h$ is a $FL$-projectable function and
that $U \in \ker FL_*$, then $U(i_{X_h} dE_L) = U \circ FL^* \circ
Y_h (H)= 0$, so
$$
0 = U(C_h) = U(i_{X_h} i_\Ga \om_L) = -\L_U (i_{X_h} i_\Ga \om_L)
= -i_\Ga \L_U i_{X_h} \om_L - i_{[U,\Ga]} i_{X_h} \om_L\ .
$$
Now, since $U \in \ker FL_*$, then $U$ is $FL$-related to 0,
therefore $\L_U \circ FL^* = 0$, so
$$
\L_U i_{X_h} \om_L = \L_U i_{X_h} FL^* \om_0
= \L_U \circ FL^* \circ i_{Y_h} \om_0 = 0\ .
$$
We conclude that
$$
i_{[U,\Ga]} i_{X_h} \om_L = 0  \sepword{for all} U \in \ker FL_*\ .
\eqlabel\ttwo
$$
On the other hand, since $S(X)$ is $\tau$-vertical, \eq\twenty\
gives $S(X)= S([S(X),\Ga])$, therefore $V:=  X - [S(X),\Ga]$ is also
$\tau$-vertical.  Now, for $X\in \X(TM)$ we can write
$$
i_X i_{X_h} \om_L =  i_{[S(X),\Ga]} i_{X_h} \om_L
+ i_V i_{X_h} \om_L\ .
\eqlabel\tthree
$$
Moreover, when $X\in \Mg$, $S(X) \in \ker FL_* = \Vker \om_L$
\cite{\CarinenaLoRo, Prop. 3}, then by \eq\ttwo\ the first
term of \eq\tthree\ vanishes, while the second one vanishes by
Lemma 4.2 and Proposition 4.4, hence $X_h \in \Mg^\perp = \ker \om_L$.
\qed

Thus, the Lagrangian dynamical constraints are exactly those
that are $FL$-projectable, while the non-projectable ones are
associated to the SODE conditions.  Moreover, this partition of
the Lagrangian constraints in two groups can also be explained
in terms of the classification of the Hamiltonian constraints:

\proclaim Theorem 4.6.
Let $h\in \Coo(T^*M)$ be a Hamiltonian constraint, $h$ is first
class if, and only if, $C_h$ is $FL$-projectable.

\Proof
If $h$ is first class  $Y_h$ is tangent to $\Im FL$, in other
words $FL^* \circ Y_h = 0$. Then
$$
\om_L(X_h,X) = \hat \om_0(X_h \circ FL^*,X \circ FL^*)
= \hat \om_0(FL^* \circ Y_h,X \circ FL^*) =0.
$$
Thus, $X_h \in \ker \om_L$, so, by Theorem 4.5, $C_h$ is
$FL$-projectable.

On the other hand, if $h$ is second class there exists another
constraint $k$ such that
$$
\leqalignno{
0 &\ne FL^*\cor{h}{k} = FL^*\om_0(Y_h,Y_k)
= \hat \om_0(FL^* \circ Y_h, FL^* \circ Y_k)
& (\ecnum) \cr
&= \hat \om_0(X_h \circ FL^* , X_k \circ FL^*)
=\om_L(X_h , X_k).  \cr}
$$
Hence $X_h \notin \ker \om_L$, and again the previous theorem
implies that $C_h$ is not $FL$-projectable.  \qed

To finish, we point out that our algebraic approach allow us
to generalize all the results to the graded context,
without changing a single word in our arguments.

\bigskip

\noindent {\bf Acknowledgements}

\smallskip

JFC acknowledges partial financial support from  Spanish DGI under project
BFM-2000-1066-C03-01.
HF thanks the Vicerrector\1a de Investigaci\'on de la
Universidad de Costa Rica for support.

\bigskip

\noindent {\bf References}

\frenchspacing

\bigskip

\refno\batlleetal.
  C. Batlle, J. Gomis, J.M. Pons and N. Rom\'an-Roy,
{\it Equivalence between the lagrangian and hamiltonian formalism
for constrained systems},
J.~Math. Phys. {\bf 27} (1986) 2953--2962.

\refno\berg.
P.G.   Bergmann and  I. Goldberg,
{\it Transformations in phase space and Dirac brackets},
Phys. Rev. {\bf 98} (1955) 531--38.

\refno\CarinenaFer.
J.F. Cari\~nena and  J. Fern\'andez--N\'u\~nez,
{\it Geometric theory of time--dependent singular Lagrangians},
Fortschr. Phys. {\bf41} (1993) 517--552.

\refno\Hector.
J.F. Cari\~nena and H. Figueroa, {\it A geometrical version of Noether's
theorem in supermechanics}, Rep. Math. Phys.
{\bf34} (1994) 277--303.

\refno\Hectoriii.
J.~F. Cari\~nena and H. Figueroa, {\it Hamiltonian versus
Lagrangian formulation of supermechanics},
J. Phys. A: Math. Gen.  {\bf 30} (1997) 2705--2724.

\refno\Hectoriv.
   J.~F. Cari\~nena and H. Figueroa, {\it Geometric formulation of
higher order Lagrangian systems in supermechanics},
Acta Appl. Math. {\bf 51} (1998) 25--58.

\refno\Hectorv.
J.~F. Cari\~nena and H. Figueroa, {\it Recursion operators and
constants of motion in supermechanics}, J. Diff. Geom. and Appl.
{\bf 10} (1999) 191--202.

\refno\CarinenaLopez.
J.F. Cari\~nena and C. L\'opez, {\it The Time--Evolution Operator for
Singular Lagrangians}, Lett. Math. Phys. {\bf14} (1987) 203--210.

\refno\CarinenaLoMari.
J.F. Cari\~nena, C. L\'opez and E. Mart\1nez, {\it A new approach
to the converse of Noether's theorem}, J. Phys. A: Math.  Gen.
{\bf 22} (1989) 4777--4786.

\refno\CarinenaLoMarii.
J.F. Cari\~nena, C. L\'opez and E. Mart\1nez, {\it Sections along a map
applied to Higher Order Lagrangian Mechanics. Noether's theorem},
Acta Appl. Math. {\bf25} (1991) 127--151.

\refno\CarinenaLoRo.
J.F. Cari\~nena, C. L\'opez and N. Roman--Roy, {\it Geometric study of
the connection between the Lagrangian and Hamiltonian constraints}, J.
Geom. Phys. {\bf4}  (1987) 315--334.

\refno\CrampinPirani.
M. Crampin and  F.A.E. Pirani, {\sl Applicable Differential
geometry}, London Mathematical Society Lecture Notes Series
{\bf 59}, Cambridge University Press, Cambridge, UK, 1988.

\refno\dirac.
P.A.M.  Dirac,
{\it Generalized
Hamiltonian dynamics},
  Can. J. Math. {\bf  2} (1950) 129--148.

\refno\catalanes.
  A. Echeverr\'{\i}a-Enr\'{\i}quez, J. Mar\'{\i}n-Solano, M.C. Mu\~noz-Lecanda,
and  N. Rom\'an-Roy, {\it Sections along maps in field theories.
The covariant field operators}, e-print math-ph/0103019.

\refno\tesis.
H. Figueroa, {\sl Variedades graduadas y sus aplicaciones en
supermec\'anica}, Ph. D. dissertation, Universidad de Zaragoza, July
1996, published in Publicaciones del Seminario Matem\'atico
Garc\1a de Galdeano, serie II, secci\'on 2, 55, (1997).

\refno\GraciaPons.
X. Gracia and J. M. Pons, {\it On an Evolution Operator connecting
Lagrangian and Hamiltonian Formalisms},  Lett. Math. Phys. {\bf 17}
(1989) 175--180.

\refno\GraciaPonsdos.
X. Gracia and J. M. Pons, {\it Singular Lagrangians: some geometric
structures along
the Legendre map},  J. Phys. A: Math.  Gen.
{\bf 34} (2001) 3047--70.

\refno\IbortMarin.
L.A. Ibort and J. Mar\1n-Solano, {\it Geometrical foundations of
Lagrangian supermechanics and supersymmetry}, Rep. Math.
Phys., {\bf32} (1993) 385--409.

\refno\kamimura.
K. Kamimura,
{\it Singular Lagrangian and Hamiltonian systems, generalized
canonical formalism},
  Nuovo Cim. {\bf 68 B} (1982) 33--54.

\refno\Kostant.
B. Kostant, {\it Graded manifolds, graded Lie theory and
prequantization}, in {\sl Differential Geometrical Methods in
Mathematical Physics}, Lecture Notes in Mathematics {\bf570},
Springer, Berlin, 1977.

\refno\Leites.
D. A. Le\u{\i}tes, {\it Introduction to the theory of
supermanifolds}, Russian Math. Surveys\ {\bf35}
(1980) 1--64.

\refno\Manin.
Yu.I. Manin, {\sl Gauge Field theory and Complex Geometry},
Nauka. Moscow, (1984). English transl., Springer--Verlag, New York,
(1988).

\refno\MunozRoman.
M.C. Mu\~noz and N. Roman--Roy, {\it Lagrangian theory for
presymplectic systems}, Ann. Inst. H. Poincar\'e {\bf 47}
(1992) 27--45.

\refno\PugliVino.
F. Pugliese and A.M. Vinogradov, {\it On the geometry of singular Lagrangians}, 
J. Geom. Phys.  {\bf 35} (2000) 35--55.

\refno\Sanchez.
O.A. S\'anchez--Valenzuela, {\it Differential--Geometric approach
to  supervector bundles}, Comunicaciones T\'ecnicas IIMASS--UNAM,
(Serie Naranja) {\bf457}, M\'exico, (1986).

\refno\Skinner.
R. Skinner, {\it First order equations for classical mechanics},
J. Math. Phys. {\bf24} (1983) 2581--2588.

\refno\SkinnerRusk.
R. Skinner and R. Rusk, {\it Generalized Hamiltonian dynamics I.
Formulation on $T^*Q \o+ TQ$}, J. Math. Phys. {\bf24} (1983)
2589--2594.

\refno\Vaintrob.
  A. Vaintrob, {\it Lie algebroids and homological vector
fields\/}, Russ. Math. Surv. {\bf 52} (1997) 428--429.

\bye